\documentstyle[preprint,aps]{revtex}
\begin{document}
\preprint{\begin{tabular}{c}
\hbox to\textwidth{October 1994 \hfill SNUTP 94-95}\\[-10pt]
\hbox to\textwidth{hep-ph/9410284 \hfill}\\[36pt]
\end{tabular}}
\draft
\title{Probing Supergravity Models\\
 with Indirect Experimental Signatures}
\author{Jihn~E.~Kim$^{(a),(b)}$ and Gye~T.~Park$^{(a)}$}
\address {\it $^{(a)}$Center for Theoretical Physics and
$^{(b)}$Department of Physics\\
%[-8pt]
Seoul National University, Seoul, 151-742, Korea}
%[-8pt]

%\date{\today}
\maketitle
\begin{abstract}

We explore the one-loop electroweak radiative corrections in the context of
the traditional minimal $SU(5)$ and
the string-inspired
$SU(5)\times U(1)$ supergravity models by calculating explicitly
vacuum-polarization and vertex-correction contributions to the $\epsilon_1$ and
$\epsilon_b$ parameters. We also include in this analysis the constraint from
$b\rightarrow s\gamma    $
whose inclusive branching ratio $B(b\rightarrow     s\gamma)    $ has been
actually measured very recently by CLEO.
We find that by combining these three most important indirect experimental
signatures
and using the most recent experimental values for them, $m_t\gtrsim 170 \;{\rm
GeV}    $ is excluded for $\mu>0$ in both the minimal $SU(5)$ supergravity
and the no-scale $SU(5)\times U(1)$ supergravity.
We also find that $m_t\gtrsim 175$ $(185)\;{\rm GeV}$ is excluded for any sign
of $\mu$ in the minimal ($SU(5)\times U(1)$) supergravity model.

\end{abstract}

%\narrowtext

\section{Introduction}
With the increasing accuracy of the LEP measurements, it has become extremely
important performing the precision test of the standard model (SM) and its
extensions.
A standard model fit to the latest LEP data yields the top mass,
$m_t=178\pm 11^{+18}_{-19}\;{\rm GeV}    $  \cite{Schaile}, which is in perfect
agreement with the measured top mass from CDF  \cite{CDF-top}, $m_t=174\pm
10^{+13}_{-12}\;{\rm GeV}    $. Therefore, it would be desirable for one to
narrow down
the top mass in the vicinity of the above central value in studying the
phenomenology of specific models of interest.
Although the SM is in remarkable agreement with all the known experiments,
there are a few experimental results that can be interpreted as a possible
manifestation of new physics beyond the SM. First,
$R_b$ ($\equiv{\Gamma(Z\rightarrow b\overline     b)\over{\Gamma(Z\rightarrow
hadrons)}}$) measured at LEP is in disagreement at $2\sigma$ level
with the SM predictions. Secondly, the flavor-changing radiative decay
$b\rightarrow     s\gamma$ \cite{Barger+Hewett,BG,bsgamma,others}, whose
inclusive branching ratio has been actually measured by CLEO
to be at 95\% C.~L. \cite{CLEO94},
$$1\times 10^{-4}<B(b\rightarrow     s\gamma)    <4\times 10^{-4},$$
still leaves room for new physics.
Large experimental value for $R_b$ would put rather perilously small upper
bound on $m_t$ in the SM  \cite{Schaile}. One could certainly
interpret this as a possible manifestation of new physics beyond the SM, where
at one loop the negative standard top quark contributions are cancelled to a
certain extent by the contributions from the new particles, thereby allowing
considerably larger $m_t$ than in the SM.
In fact, the minimal supersymmetric standard model (MSSM) realizes this
possibility \cite{epsbKP,wells}.
Similarly in $b\rightarrow     s\gamma$, the values of the branching ratio near
the current lower bound can be accomodated for reasonable values of $m_t$ in
the MSSM because suppression
can occur due to the additional contributions in the model.

In supergravity (SUGRA) models, radiative electroweak symmetry breaking
mechanism \cite{EWx} can be described by at most 5 parameters: the top-quark
mass ($m_t$), the ratio of Higgs
vacuum expectation values ($\tan\beta    $), and three universal
soft-supersymmetry-breaking parameters ($m_{1/2},m_0,A$) \footnote{See,
however, Ref.~\cite{non-univ} for
non-universal soft-supersymmetry breaking parameters}.
Since the entire range of sparticle mass spectrum is quite broad for the models
we consider, the large hadron collider (LHC) and the next linear collider (NLC)
are needed in order to explore all the regions of the parameter space of our
interest.
However, the present collider facilities have been successful in probing a good
part of the allowed parameter space through indirect experimental signatures.
In particular, we will concentrate here on the precision measurements at LEP
and the flavor-changing radiative decay $b\rightarrow     s\gamma$ observed by
CLEO.
We adopt the $\epsilon$-scheme \cite{ABJ,ABC} for a global analysis of the
precision data at LEP.
Among four $\epsilon$-parameters, $\epsilon_i$ $(i=1,2,3,b)$ in this scheme,
$\epsilon_b$ has been studied very recently in the context of the minimal
$SU(5)$ and the no-scale $SU(5)\times U(1)$ supergravity models \cite{epsbKP}.
In this work we expand the analysis to include the additional constraints from
$\epsilon_1$ and $b\rightarrow     s\gamma$ in the minimal $SU(5)$ and
a larger class of $SU(5)\times U(1)$ supergravity models.
We will show that by combining above three most important constraints
from indirect processes, $m_t\gtrsim 170 \;{\rm GeV}    $ is excluded for
$\mu>0$ in both the minimal $SU(5)$
and the no-scale $SU(5)\times U(1)$ SUGRA models.

\section{The minimal $SU(5)$ and $SU(5)\times U(1)$ SUGRA models}
We consider the minimal $SU(5)$ SUGRA model \cite{su5sugra} and $SU(5)\times
U(1)$ SUGRA model \cite{flippedsugra}
which can be regarded as traditional versus string-inspired unified models.
These two models both contain, at low energy, the
SM gauge symmetry as well as the particle content of the MSSM, that is, the SM
particles with two Higgs doublets and their superpartners. There are, however,
a few crucial differences between the two models which are: \\
(i) The unification groups are different, $SU(5)$ versus $SU(5)\times U(1)$.\\
(ii) The gauge coupling unification occurs at
$\sim 10^{16}\;{\rm GeV}    $ in the minimal $SU(5)$ model whereas in
$SU(5)\times U(1)$ model it occurs at the string scale $\sim 10^{18}\;{\rm GeV}
   $ \cite{LNZstring}.
In $SU(5)\times U(1)$ SUGRA, the gauge unification is delayed because of the
effects of
an additional vector-like quark doublet with a mass $\sim 10^{12}\;{\rm GeV}
$
and one additional vector-like quark singlet of charge $-1/3$ with a mass
$\sim 10^{6}\;{\rm GeV}    $.
The different heavy field content at the unification scale leads to different
constraints from proton decay. \\
(iii) The minimal $SU(5)$
SUGRA is highly constrained by proton decay while $SU(5)\times U(1)$ SUGRA is
not.\\
The above SUGRA models can be completely described by only five parameters
under a few simplifying assumptions on the values of the
soft-supersymmetric-breaking
parameters at the unification scale. That is, all three gauginos are assumed
to have a common mass $m_{1/2}$, and all squarks, sleptons, and two Higgs
scalar doublets to have a common mass $m_0$, and three trilinear scalar
couplings are taken to be identical to $A$.
The Higgs mixing parameter $\mu$ and its associated bilinear coupling
$B$ are in fact determined by imposing the radiative EW breaking condition.
All these boil down to only five parameters, $m_{1/2},m_0,A,\tan\beta$, and
$m_t$.
One can also restrict further the above 5-dimensional parameter spaces as
follows \cite{aspects}.
First, upon sampling a specific choice of ($m_{1/2},m_0,A$) at the unification
scale and ($m_t,\tan\beta    $) at the electroweak scale, the renormalization
group equations (RGE) are run from the unification scale to the electroweak
scale, where the radiative electroweak breaking condition is imposed by
minimizing the effective 1-loop Higgs potential to determine $\mu$ up to its
sign and $B$.
Here the sign of $\mu$ is given as usual \cite{Bargeretal}, and differs from
that of Ref.~\cite{bsgamma,epsbKP}; {\it i.e.}    , we define $\mu$ by $W_\mu
=\mu H_1 H_2$.
We also impose consistency constraints such as perturbative unification and the
naturalness bound of $m_{\tilde g}\lesssim 1\,{\rm TeV}    $.
Finally, all the known experimental bounds on the sparticle masses are imposed
\footnote{We use the following experimental lower bounds on the sparticle
masses in GeV in the order of gluino, squarks, lighter stop, sleptons, and
lighter chargino: $m_{\tilde g}\gtrsim 150$, $m_{\tilde q}\gtrsim 100$,
$m_{{\tilde{t}}_1}\gtrsim 45$, $m_{\tilde l}\gtrsim 43$,
$m_{\chi^\pm_1}\gtrsim 45$.}.
This prodedure yields the restricted parameter spaces for the two models.

Further reduction in the number of input parameters in $SU(5)\times U(1)$ SUGRA
is made possible because in specific string-inpired models for
($m_{1/2},m_0,A$) at the unification scale these three parameters are computed
in terms of just one of them \cite{IL}. One obtains $m_0=A=0$ in the {\em
no-scale} model and
$m_0=\frac{1}{\sqrt{3}}m_{1/2}$, $A=-m_{1/2}$ in the {\em dilaton} model
\footnote{Note, however, that one loop correction changes this relation
significantly \cite{cosCKN}.}.

The low energy predictions for the sparticle mass spectra are quite different
in the two SUGRA models mainly due to the different pattern of supersymmetry
radiative breaking.
In the minimal $SU(5)$ SUGRA model, all the squarks except the lighter stop and
all the Higgs except the lighter neutral Higgs are quite heavy ($\gtrsim$ a few
hundred GeV) whereas in the $SU(5)\times U(1)$ SUGRA model they can be quite
light.
This difference leads to strikingly different phenomenology in the two models,
for example in the flavor changing radiative decay $b\rightarrow s\gamma    $
\cite{bsgamma}.

\section{Constraints from the EW radiative corrections and the flavor changing
radiative decay}
Parametrizing the electroweak vacuum
polarization corrections with three parameters can be understood as follows.
It can be shown, by
expanding the vacuum polarization tensors to order $q^2$, that one obtains
three independent physical parameters. Alternatively, one can show that upon
symmetry breaking three additional terms appear in the effective lagrangian
 \cite{efflagr}.
Among several schemes to parametrize the corrections
\cite{efflagr,Kennedy,PT,AB},
in the $(S,T,U)$ scheme  \cite{PT}, the deviations of the model
predictions from the SM predictions (with fixed SM values for $m_t,m_H$)
are considered as the effects from ``new physics". This scheme is valid only up
to the lowest order in $q^2$, and is therefore not applicable to a theory with
light new particles comparable to $M_Z$. In the $\epsilon$-scheme
\cite{ABJ,ABC}, on
the other hand, the model predictions are absolute and also valid up to higher
orders in $q^2$, and therefore this scheme is more applicable to the
electroweak
precision tests of the MSSM  \cite{BFC} and a class of supergravity models
 \cite{ewcorr}.

There are two different $\epsilon$-schemes. The original scheme \cite{ABJ} was
considered in one of author's previous analyses  \cite{ewcorr,bsg-eps}, where
$\epsilon_{1,2,3}$ are defined by a basic set of observables $\Gamma_{l},
A^{l}_{FB}$ and $M_W/M_Z$.
Due to the large $m_t$-dependent vertex corrections to $\Gamma_b$, the
$\epsilon_{1,2,3}$ parameters   and $\Gamma_b$ can be correlated only for a
fixed value of $m_t$. Therefore, $\Gamma_{tot}$, $\Gamma_{hadron}$ and
$\Gamma_b$ were not included  in Ref.~\cite{ABJ}. However, in the new
$\epsilon$-scheme, introduced recently in Ref.~\cite{ABC}, the above
difficulties are overcome by introducing a new parameter $\epsilon_b$ to encode
the $Z\rightarrow b\overline b    $ vertex corrections. The four $\epsilon$'s
are now defined by an
enlarged set of $\Gamma_{l}$, $\Gamma_{b}$, $A^{l}_{FB}$ and $M_W/M_Z$ without
even specifying $m_t$.
This new scheme was adopted in a previous analysis by one of us (G.P.) in the
context of the $SU(5)\times U(1)$ SUGRA models \cite{LNPZepsb}.
In this work we use this new $\epsilon$-scheme.
As is well known, the SM contribution to $\epsilon_1$ depends quadratically
on $m_t$ but only logarithmically on the SM Higgs boson mass ($m_H$). Therefore
upper bounds on $m_t$  have a non-negligible $m_H$
dependence and become around $20\;{\rm GeV}    $ lower when going from
$m_H=1\;{\rm TeV}    $
to $m_H=100\;{\rm GeV}    $. It is also known in the MSSM that
the largest supersymmetric contributions to $\epsilon_1$ are expected to
arise from the $\tilde t$-$\tilde b$ sector, and in the limiting case of a very
light stop the contribution is comparable to that of the $t$-$b$ sector. The
remaining squark, slepton, chargino, neutralino, and Higgs sectors all
typically contribute considerably less. For increasing sparticle masses, the
heavy sector of the theory decouples, and only SM effects  with a {\it light}
Higgs boson survive. However, for a
light chargino ($m_{\chi^\pm_1}\rightarrow    {1\over2}M_Z$), a
$Z$-wavefunction
renormalization threshold effect coming from Z-vacuum polarization diagram with
the lighter chargino in the loop
 can introduce a substantial $q^2$-dependence
in the calculation  \cite{BFC}.
This results in a weaker upper bound on $m_t$ than in the SM.
The complete vacuum polarization contributions from the Higgs sector, the
supersymmetric chargino-neutralino and sfermion sectors, and also the
corresponding contributions in the SM have been included in our calculations
 \cite{ewcorr}. However, the supersymmetric contributions to the non-oblique
corrections except in $\epsilon_b    $ have been neglected.

Following Ref.~\cite{ABC}, $\epsilon_b    $ is defined from $\Gamma_b$, the
inclusive
partial width for $Z\rightarrow b\overline b    $, as
\begin{equation}
\epsilon_b    ={g^b_A\over{g^l_A}}-1
\end{equation}
where $g^b_A$ $(g^l_A)$ is the axial-vector coupling of $Z$ to $b$ $(l)$.
In the SM, the diagrams for $\epsilon_b    $  involve top quarks and
$W^\pm$ bosons  \cite{RbSM}, and the contribution to $\epsilon_b    $ depends
quadratically on $m_t$ ($\epsilon_b    =-G_F m_t^2/4\sqrt {2}\pi^2 + \cdots$).
In supersymmetric models there are additional diagrams
involving Higgs bosons and supersymmetric particles. The charged Higgs
contributions have been calculated in Refs.~ \cite{Denner,epsb2HD} in
the context of a non-supersymmetric two Higgs doublet model, and the
contributions involving supersymmetric particles in Refs.~ \cite{BF,Rb2HD}.
The main features of the additional supersymmetric contributions are: (i) a
negative contribution
from charged Higgs--top exchange which grows as $m^2_t/\tan^{2}\beta$ for
$\tan\beta\ll{m_t\over{m_b}}$; (ii) a positive contribution from chargino-stop
exchange which in this case grows as $m^2_t/\sin^{2}\beta$; and (iii) a
contribution from neutralino(neutral Higgs)--bottom exchange which grows as
$m^2_b\tan^{2}\beta$ and is negligible except for large values of $\tan\beta$
({\it i.e.}    , $\tan\beta\gtrsim{m_t\over{m_b}}$) (the contribution (iii) has
been
neglected in our analysis).

In the MSSM, $b\rightarrow s\gamma    $ decay receives significant
contributions from penguin diagrams with $W^\pm-t$ loop, $H^\pm-t$ loop
\cite{HCbsg} and the $\chi^\pm_{1,2}-\tilde t_{1,2}$ loop \cite{Bertolini}.
The expression used for $B(b\rightarrow     s\gamma)    $ in the leading
logarithmic (LL) calculations is given by \cite{GSW}
\begin{equation}
{B(b\rightarrow     s\gamma)\over B(b\rightarrow
ce\bar\nu)}={6\alpha\over\pi}
{\left[\eta^{16/23}A_\gamma
+{8\over3}(\eta^{14/23}-\eta^{16/23})A_g+C\right]^2\over
I(m_c/m_b)\left[1-{2\over{3\pi}}\alpha_s(m_b)f(m_c/m_b)\right]},\label{bsg}
\end{equation}
where $\eta=\alpha_s(M_W)/\alpha_s(m_b)$, $I$ is the phase-space factor
$I(x)=1-8x^2+8x^6-x^8-24x^4\ln x$, and $f(m_c/m_b)=2.41$ the QCD
correction factor for the semileptonic decay.
$C$ represents the leading-order QCD
corrections to the $b\rightarrow s\gamma    $ amplitude when evaluated at the
$\mu=m_b$ scale
\cite{GSW}.
We use the 3-loop expressions for $\alpha_s$ and choose $\Lambda_{QCD}$ to
obtain $\alpha_s(M_Z)$ consistent with the recent measurements at LEP.
In our computations we have used: $\alpha_s(M_Z)=0.118$, $ B(b\rightarrow
ce\bar\nu)=10.7\%$, $m_b=4.8\;{\rm GeV}    $, and
$m_c/m_b=0.3$. The $A_\gamma,A_g$ are the
coefficients of the effective $bs\gamma$ and $bsg$ penguin operators
evaluated at the scale $M_W$. Their simplified expressions are given in
Ref.~\cite{BG}
in the justifiable limit of negligible gluino and neutralino contributions
\cite{Bertolini} and degenerate squarks, except for the $\tilde t_{1,2}$ which
are significantly split by $m_t$.
Regarding large uncertainties in the LL QCD corrections, which is mainly due to
the choice of renormalization scale $\mu$ and is estimated to be $\approx
25\%$, it has been recently demonstrated by Buras {\it et al.} in
Ref.~\cite{burasetal}
that the significant $\mu$ dependence in the LL result can in fact be reduced
considerably by including next-to-leading logarithmic (NLL) corrections, which
however, involves very complicated calculations of three-loop mixings
between cetain effective operators and therefore have not been completed yet.

\section{Results and discussion}
In Figure 1 we present our numerical results for $\epsilon_1$ versus
$\epsilon_b$ in the two $SU(5)\times U(1)$ SUGRA
models. Similar analysis, within the context of the infrared fixed point
solution of the top quark mass in the MSSM, was recently performed in
Ref.~\cite{carena}.
$\alpha_S(M_Z)=0.118$ and $m_b=4.8\;{\rm GeV}    $ are used throughout the
numerical calculations.
We use in the figure the following experimental values for $\epsilon_1$ and
$\epsilon_b    $,
$$\epsilon_1^{exp}=(3.5\pm 1.8)\times 10^{-3}, \qquad \epsilon_b^{exp}=(0.9\pm
4.2)\times 10^{-3},$$ determined from the latest $\epsilon$- analysis using the
LEP and SLC data in Ref.~\cite{Altarelli94}.
In the figure points between the two horizontal lines are allowed by the
$\epsilon_1$ constraint at the 90\% C.~L. while the arrow points into the
region allowed by
the $\epsilon_b$ constraint at the 90\% C.~L.
The values of $m_t$ are as indicated.
The combined constraint is stronger for $\mu>0$, excluding $m_t\gtrsim
180\;{\rm GeV}    $ at the 90\% C.~L.
The significant drop in $\epsilon_1$ comes from the threshold effect of
Z-wavefunction renormalization as discussed in the previous section.
The current experimental values for $\epsilon_{1,b}$ prefer light but not too
light chargino: for $m_t=170\;{\rm GeV}    $, in the no-scale (dilaton) model,
$$50\;{\rm GeV}    \lesssim m_{\chi^\pm_1}\lesssim 70 \;(60)\;{\rm GeV}     ,
\qquad \mu>0$$
$$50\;{\rm GeV}    \lesssim m_{\chi^\pm_1}\lesssim 150 \;(140)\;{\rm GeV}     ,
\qquad \mu<0$$
where $m_{\chi^\pm_1}\lesssim 50\;{\rm GeV}    $ is disfavored by the
$\epsilon_1$ constraint \footnote{Our calculation of $\epsilon_1$ near the
threshold ($m_{\chi^\pm_1}\rightarrow    {1\over2}M_Z$) becomes unstable and
loses its credibility.}.

Since $\epsilon_{1,b}$ constrains the models the most for $m_t=170\;{\rm GeV}
 $,
we show in Figure 2
the model predictions for $B(b\rightarrow     s\gamma)    $ versus $\epsilon_b
  $ for $m_t=170\;{\rm GeV}    $
to see if $b\rightarrow s\gamma    $ provides an additional constraint.
It is very interesting to see that the combined constraint of $b\rightarrow
s\gamma    $ and $\epsilon_b    $ can in fact exclude $m_t=170\;{\rm GeV}    $
for $\mu>0$
in the no-scale model.
Therefore, combining $\epsilon_{1,b}$ and $b\rightarrow s\gamma    $ allows one
to exclude
$m_t\gtrsim 170\;{\rm GeV}    $ for $\mu>0$ altogether in the no-scale model.
Similarly, in the dilaton model, $m_t\gtrsim 170\;{\rm GeV}    $ for $\mu>0$
is almost excluded.
The large suppression in $B(b\rightarrow     s\gamma)    $ for $\mu<0$ in these
models is worth
further explanation. As first noticed in Ref. \cite{bsgamma},
what happens is that in Eq.~(\ref{bsg}), the $A_\gamma$ term nearly
cancels against the QCD correction factor $C$; the $A_g$ contribution is small.
The $A_\gamma$ amplitude receives three contributions: from the $W^\pm$-$t$
loop, from the $H^\pm$-$t$ loop, and from the
$\chi^\pm_{1,2}$-${\tilde t}_{1,2}$ loop. The first two contributions are
always negative \cite{GSW}, whereas the last one can have either sign, making
it possible having cancellations among three contributions.

In Figure 3-4, we also present the model predictions in the minimal $SU(5)$
SUGRA model for $\epsilon_1$ versus
$\epsilon_b$ (top row) and for $B(b\rightarrow     s\gamma)    $ versus
$\epsilon_b$ (bottom row) for $m_t=160, 175, 190\;{\rm GeV}    $ (Fig.~3) and
for $m_t=170\;{\rm GeV}    $ (Fig.~4).
As can be seen in the figure, the additional constraint from $B(b\rightarrow
 s\gamma)    $ here is rather mild as compared to the one in the $SU(5)\times
U(1)$ models. However, $\epsilon_b$ constraint turns out to be the strongest of
all,
excluding $m_t\gtrsim 175\;{\rm GeV}    $ at the 90\% C.~L.
For $m_t=170\;{\rm GeV}    $, unlike in the $SU(5)\times U(1)$ SUGRA, $\mu>0$
is excluded
by the $\epsilon_1-\epsilon_b    $ constraint alone, thereby excluding
$m_t\gtrsim 170\;{\rm GeV}    $
with $\mu>0$ in the minimal $SU(5)$ SUGRA model.
For $\mu>0$, $\epsilon_b    $ prefers light chargino, $m_{\chi^\pm_1}\lesssim
110\;{\rm GeV}    $.
In comparison with the results of $\epsilon_{1,b}$ in the $SU(5)\times U(1)$
SUGRA models (Fig.~1), especially the rise in $\epsilon_b    $ in the minimal
$SU(5)$ SUGRA model is less pronounced.
This is mainly due to the fact that
the stop mass, which is responsible for the rise in $\epsilon_b    $ when it
gets light, in fact scales with the chargino mass in the $SU(5)\times U(1)$
SUGRA model
whereas it does not in the minimal $SU(5)$ SUGRA model.
Therefore, the stop mass and the chargino mass become lighter simultaneoulsly
only in the $SU(5)\times U(1)$ SUGRA models, making
the light chargino effect in $\epsilon_b$ more manifest
than in the minimal $SU(5)$ SUGRA model.
This difference of course leads to different $\epsilon_b    $-deduced $m_t$
bounds in these models \cite{epsbKP}.

We would like to comment on the possible direct experimental signatures
at the present and future colliders in the models considered here.
At the Tevatron, with the estimated sensitivity range by the end of the
on-going Run IB the trilepton signals from chargino-neutralino production can
probe $m_{\chi^\pm_1}\lesssim 80-90\;{\rm GeV}    $ only in the dilaton model
but not in the no-scale model.
The preferred signal at LEPII for the chargino pair production is so-called
``mixed-mode'' $(1\ell +2j)$, with which one may be able to probe
$m_{\chi^\pm_1}\lesssim 96\;{\rm GeV}    $ only in the dilaton model
\cite{easpects}.

\section{Conclusions}

We have computed the one-loop electroweak radiative corrections
in terms of $\epsilon_1$ and $\epsilon_b$ in the context of the traditional
minimal $SU(5)$ and the string-inspired
$SU(5)\times U(1)$ supergravity models. We have also considered
the strongest constraint from the radiative flavor-changing decay
$b\rightarrow s\gamma    $ as prompted by the recent measurement of CLEO on the
inclusive branching ratio $B(b\rightarrow     s\gamma)    $.
We use the latest experimental values for $\epsilon_1$, $\epsilon_b$
and $B(b\rightarrow     s\gamma)    $
in order to constrain the models.
We find that combining constraints from $\epsilon_{1,b}$ and $b\rightarrow
s\gamma    $ allows one to exclude
$m_t\gtrsim 170\;{\rm GeV}    $ for $\mu>0$ altogether in both the minimal
$SU(5)$ SUGRA
and the no-scale $SU(5)\times U(1)$ SUGRA.
We also find that $m_t\gtrsim 175$ $(185)\;{\rm GeV}$ is excluded for any sign
of $\mu$ in the minimal ($SU(5)\times U(1)$) SUGRA model.

Our results on $B(b\rightarrow     s\gamma)    $ is subject to change due to a
large uncertainty
in QCD corrections. One could perhaps draw more precise conclusions when
QCD corrections become better known.

\acknowledgments
We would like to thank the Institute for Elementary Particle Physics Research,
University of Wisconsin for the use of their computer facilities.
This work is supported in part by the Korea Science and Engineering
Foundation through Center for Theoretical Physics, Seoul
National University (JEK,GTP), KOSEF--DFG Collaboration Program (JEK),
the Basic Science Research Institute Program, Ministry of
Education, 1994, BSRI-94-2418 (JEK).

\begin{figure}
\caption{The correlated predictions for $\epsilon_1$ and
$\epsilon_b$ in $10^{-3}$ in the no-scale (top row) and the dilaton (bottom
row)
$SU(5)\times U(1)$ supergravity model.
Points between the two horizontal lines are allowed by the $\epsilon_1$
constraint at the 90\% C.~L. while the arrow points into the region allowed by
the $\epsilon_b$ constraint at the 90\% C.~L.
The values of $m_t$ are as indicated.}
\label{bsgeps1b1}
\end{figure}
\begin{figure}
\caption{The correlated predictions for $B(b\rightarrow     s\gamma)    $ and
$\epsilon_b$ in the no-scale (top row) and the dilaton (bottom row)
$SU(5)\times U(1)$ supergravity model for $m_t=170\;{\rm GeV}    $.
Points between the two horizontal lines are allowed by the $b\rightarrow
s\gamma    $ constraint at the 95\% C.~L. while the arrow points into the
region allowed by
the $\epsilon_b$ constraint at the 90\% C.~L.}
\label{bsgeps1b2}
\end{figure}
\begin{figure}
\caption{The correlated predictions in the minimal $SU(5)$ supergravity model
for $\epsilon_1$ versus
$\epsilon_b$ in $10^{-3}$ (top row) and for $B(b\rightarrow     s\gamma)    $
versus
$\epsilon_b$ (bottom row).
Points between the two horizontal lines are allowed by the $\epsilon_1$ or
$b\rightarrow s\gamma    $ constraint while the arrow points into the region
allowed by
the $\epsilon_b$ constraint at the 90\% C.~L.
The values of $m_t$ are as indicated.}
\label{bsgeps1b3}
\end{figure}
\begin{figure}
\caption{Same as in the Fig.~3 except that $m_t=170\;{\rm GeV}    $ is used.}
\label{bsgeps1b4}
\end{figure}
%\begin{figure}
%\end{figure}

% \begin{table}
% \caption{}
% \label{}
% \end{tabular}
% \end{table}
%\newpage

\end{document}